\documentclass[11pt,a4paper,reqno,normal,draft,english]{ijocta}
\usepackage[final]{graphicx}
\usepackage[margin=2cm]{geometry}
\usepackage[font=small,labelfont=bf,tableposition=top]{caption}
\usepackage[font=footnotesize]{subcaption}
\usepackage{multicol}
\usepackage{babel}
\usepackage{amssymb,amsmath,amsfonts}
\usepackage{multirow}
\usepackage{blindtext} 
\usepackage{adjustbox}
\usepackage{hyperref}
\usepackage[utf8]{inputenc}
\DeclareUnicodeCharacter{2016}{-}% support older LaTeX versions
\usepackage[numbers]{natbib}
\usepackage{url}

\begin{document}

% \begin{tabular}{ll}
% {\hspace{7pt} \scriptsize An International Journal of Optimization} & \hspace{6cm} \multirow{2}{*}{%
% \includegraphics[height=1.9cm, width=2.9cm]{ijocta_logo}} \\ 
% {\hspace{7pt} \scriptsize and Control: Theories }{$\And$}{\scriptsize Applications} & \\ 
% {\hspace{7pt} \scriptsize Vol.3, No.1, pp. (2020) \copyright\ IJOCTA} & \\ 
% {\hspace{7pt} \scriptsize ISSN:2146-0957  eISSN:2146-5703} &\\ 
% {\hspace{7pt} \scriptsize http://www.ijocta.com}
% \end{tabular}

\vspace{1.42cm}

\title[\small{\textit{Appointment scheduling model in healthcare using clustering algorithms}}]{\textbf{Appointment scheduling model in healthcare using clustering algorithms}}
\author[\small{\textit{N. Yousefi, F. Hasankhani, M. Kiani\\/Vol., No., pp. (2020) \copyright IJOCTA}}]{\normalsize Niloofar Yousefi$^a$, Farhad Hasankhani$^b$*, Mahsa Kiani$^b$, Nooshin Yousefi$^c$ \vspace{0.58cm} %please don't remove \vspace{.3cm}
\\ \small
$^a$Department of Computer Science,University of Arkansas at Little Rock, USA \\
Email: n.yousefi@ualr.edu \vspace{0.1cm} \\
$^b$Department of Industrial Engineering, Clemson University, USA \\
$^c$Department of Industrial and Systems Engineering, Rutgers University, USA \\
}

\begin{abstract}

In this study, we provided a scheduling procedure which is a combination of machine learning and mathematical programming that minimizes the waiting time of higher priority outpatients. Outpatients who request for appointment in healthcare facilities have different priorities. Determining the priority of outpatients and allocating the capacity based on the priority classes are important concepts that have to be considered in the scheduling of outpatients. In this study, two stages are defined for scheduling an incoming outpatient. In the initial stage, we employed and evaluated four distinct clustering techniques; K-means clustering, agglomerative hierarchical clustering, DBSCAN, and OPTICS clustering to classify outpatients into priority classes and suggested the best pattern to cluster the outpatients. In the second stage, we modeled the scheduling problem as a Markov Decision Process (MDP) problem since the arrivals are uncertain and the decisions are taken at the end of each day after observing total requests. Due to the curse of dimensionality, we used the fluid approximation method to estimate the optimal solution of the MDP. our methodology is employed in a data set of Shaheed Rajaei Medical and Research Center, and we represented how our model works in prioritizing and scheduling outpatients.

\vspace{7pt}

\vspace{2pt}
\noindent
%\textbf{AMS Classification:} Find your AMS Code from %\url{http://www.ams.org/mathscinet/msc/msc2010.html}
\end{abstract}

\maketitle

\section{Introduction}

\noindent Nowadays, patients in many healthcare facilities face the problem of long waiting times. Waiting times in healthcare clinics are categorized into ``indirect waiting time'' and ``direct waiting time''. Indirect waiting time is mostly expressed in days and is defined as the number of days between appointment request day and appointment day. Direct waiting time is defined as the time that a patient spends in a clinic to see a doctor.
Offering appointments with low indirect waiting time is one of the healthcare managers' issues. Long indirect waiting times may bring medical impacts, especially for multi-comorbidity and higher-priority patients. Long indirect waiting times also increase the no-show probability of patients which decreases the utilization of the healthcare facility.
Prioritization of patients based on their comorbidities and characteristics and deciding which one should get a sooner appointment is not a simple problem. Many factors play important roles in determining the level of urgency of a patient. Machine learning methods provide a decision-making tool for grouping patients into different priority classes which is more accurate than a human diagnosis. Considering patients' backgrounds and environments for clustering patients are important issues that humans may ignore. Therefore, having a tool to find a pattern for patients' priority, considering patients' histories and environmental factors helps the healthcare facilities to come up with a more accurate priority diagnosis and following that a better scheduling process. 

\vspace{1mm}

The priority of patients is an important issue that has to be considered in their scheduling. There are several methods to set a priority rule for the patients before appointment scheduling. However, most of these priority rules are deterministic and fixed based on a few parameters
\cite{godin2010agent}. 
There are different parameters and patient’s characteristics that may affect their priority levels which are not considered in the previous studies. Family history, gender, age, etc., are some features that are not considered in calculating the priority level of patients in the previous studies. In the proposed study, a machine learning method is used to consider more factors to categorize the outpatients into different priority classes. The combination of some negligible features may have a large impact on the health of the patients which is ignored in the previous research, but is considered in the machine learning methods to find the priority class of each patient. 

\vspace{1mm}

In this study, we propose a scheduling model in which we categorize outpatients into priority classes based on their comorbidities. We compared four machine learning methods; K-means clustering, agglomerative hierarchical clustering, DBSCAN, and OPTICS clustering to prioritize outpatients. Then, we schedule outpatients based on the determined priority classes within a planning horizon. We used a Markov Decision Process (MDP) model to represent the scheduling process. The reason that we use a MDP model is that the arrivals are not certain and follow Poisson processes and the decisions are made at the end of each day after observing the number of requests on that day and the number of available appointments in future days. It means that the scheduler observes all the arrivals on each day and decides about them at the end of that day. Since our MDP problem is a large-scale model, we applied the fluid approximation method to approximate the MDP solution. Since it may be difficult for the healthcare facility to solve the MDP model every day to decide regarding the scheduling of outpatients, we propose a benchmark policy that is easy to implement. This benchmark policy is obtained based on the optimal policy that is the result of solving the MDP model. Therefore, at the end of each day, the scheduler just needs to classify outpatients, and then schedule them based on the benchmark policy. Figure 1 represents a summary of our procedure in this paper for obtaining a benchmark policy for scheduling of multiple classes of outpatients. Our study is the first study in the literature that combines both machine learning methods and MDP modeling to optimize the scheduling process to decrease indirect waiting time of higher priority outpatients in receiving appointments.

\medskip
\noindent
\begin{minipage}{\linewidth}
\centering
\includegraphics[width=10cm,height=10cm]{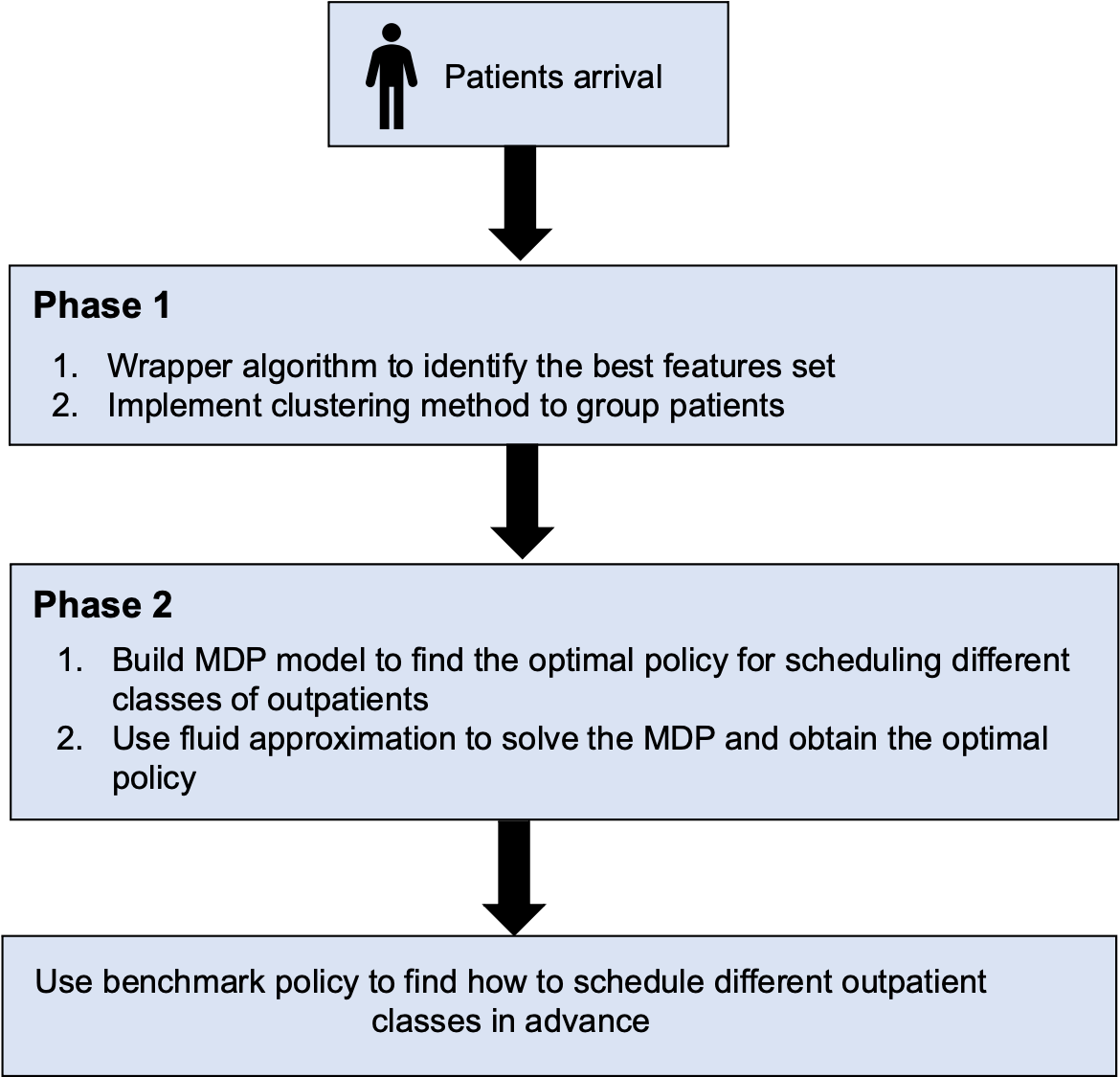}
\captionof{figure}{Summary of the proposed method}\label{fig:f5}
\end{minipage}
\medskip
\noindent

In Section 2, we discuss the uniqueness of our study concerning prior literature. Section 3.1 explains how we applied K-means clustering and agglomerative hierarchical clustering, DBSCAN, and OPTICS clustering to prioritize outpatients. The MDP model and our solution approach to solving it are fully explained in Section 3.2. We tested our model in a data set of Shaheed Rajaei Medical and Research Center in Iran and the results are presented in Section 4. Finally, Section 5 concludes this study. 

\section{Related Works}

\noindent There has been significant research on the advantages of machine learning development in healthcare systems. Reducing medical costs, improving disease diagnostics, hospital appointment scheduling, and medical research initiatives are the possible impacts of machine learning techniques on improving healthcare systems. 
\subsection{Machine learning literature review}
Machine learning is a collection of data analytical techniques programmed to learn patterns from data sets. Using mathematical rules and statistical assumptions, machine learning methods develop a pattern/model among the features of the data sets \cite{sadrfaridpour2016algebraic,sadrfaridpour2019engineering,Niloofar2023Toxicityand}. There are two main learning techniques: supervised learning and unsupervised learning. Clustering methods are the most common unsupervised learning methods. Through the training steps of the machine learning methods, the optimal model's parameters are found by calculating the errors and evaluating the model's performance through some back-and-forth steps. Then, using the optimal parameters, the model can be used for any new data set \cite{Niloofar2023Comparing}. 

\vspace{1mm}

Recently, there have been significant developments and attention on machine learning methods in different industries. Healthcare is one of the areas that significantly benefit from the development of machine learning techniques. Machine learning has the potential to help both patients and providers in terms of better care and lower costs. \cite{huang2014toward,yousefi2021master} developed a machine learning model for predicting the diagnosis of depression up to one year in advance. \cite{pendharkar2014machine} compared three different machine learning prediction methods for predicting patient's length of stay in Pennsylvania Federal and Specialty hospitals. \cite{samorani2015outpatient} used machine learning methods to obtain a show up probability for individual appointments and proposed a hospital scheduling appointment model using the show up probability of each appointment. \cite{podgorelec1997genetic} developed a model for patient scheduling using genetic algorithm and machine learning.

\vspace{1mm}

There have been several studies on the classification and clustering of patients based on specific diseases. \cite{nieuwenhuis2012classification} proposed a model for classifying schizophrenia patients based on their brain Magnetic Resonance Imaging (MRI) scans using the Support Vector Machine (SVM) method. \cite{jena2015distributed} compared different classification algorithms to predict chronic-kidney-disease. \cite{manimekalai2016prediction} compared different machine learning classifiers for predicting heart disease, and represented that SVM classifier with genetic algorithm has better prediction accuracy. \cite{chaurasia2017data} concentrated on the detection of breast cancer using different machine learning methods such as RepTree and Radial Basis Function (RBF) Network.

\vspace{1mm}

\cite{shouman2012integrating} integrated decision tree and K-means clustering to predict heart disease on Cleveland Clinic Foundation Heart disease data set. \cite{adegunsoye2018phenotypic} used the clustering method to identify four patient groups with interstitial lung disease, and showed that grouping patients could improve the efficacy of therapeutic interventions in future clinical trials. 

\vspace{1mm}

Clustering patients into different priority classes immediately after their arrival could help the healthcare facility schedule patients in a better way. 
Reviewing and analyzing large amounts of data gathered from clinical trials could improve healthcare systems in terms of disease diagnostics, patient appointment scheduling, etc. For instance, if a patient comes in with a particular case of the flu, a physician in the past would rely on what he or she knew about the flu in general or what other doctors in the area knew; while using big data analytics and machine learning methods provides a more accurate decision making tool for this diagnosis which depends on various factors in patients background that a human could ignore in his/her discernment. This fact could also be true for a hospital receptionist who schedules the patients' appointments. The priority of a patient to visit a doctor relies on different factors that a human could not pay attention to. Machine learning methods could be a significant help in classifying patients based on all their backgrounds and environments into different priority classes. Using machine learning and data analytic methods brings a more accurate way for district patients to be scheduled to visit a doctor. In this study, we used two methods to cluster the outpatients into priority classes to be scheduled upon their arrival. 

\vspace{1mm}

\subsection{Outpatient scheduling literature review}
Scheduling is the process of controlling and allocating resources to optimize the total workload of a process. There has been a significant number of scheduling problems in different areas of healthcare, manufacturing, transportation, etc. \cite{hasankhani2019time, kiani2020evaluating}. 
Scheduling of outpatients is the subject of many previous studies in healthcare systems literature. \cite{magerlein1978surgical}  provided a comprehensive review of the outpatients scheduling studies. \cite{magerlein1978surgical} classified the studies into two main groups: `` scheduling'' in which patients are scheduled in advance, and ``allocation scheduling'' in which available patients are scheduled on the service day. Our scheduling model is advanced.
\cite{patrick2008dynamic} introduced an advanced dynamic scheduling system. The decisions are made at the end of each day, and the outpatients who did not receive an appointment join the next day's waiting queue. \cite{chen2014sequencing} proposed an appointment model of a combination of advance and allocation scheduling. The model determines when the same day appointments should be scheduled throughout the day and how these same day appointments affect the routine appointments. \cite{patrick2012markov} introduced an MDP model and showed that a short booking window works better than doing the same day appointments in minimizing the total cost of the system due to unused capacity that allocation scheduling may cause.

\vspace{1mm}

The application of mathematical programming in outpatient scheduling has always been of interest to healthcare system researchers. The mathematical models used in outpatient studies are deterministic or stochastic. Deterministic models are mostly used in specialty clinics with deterministic service times. Most of the deterministic outpatient scheduling systems are formulated as integer or mixed-integer models \cite{ahmadi2017outpatient}. \cite{perez2011patient} used deterministic formulation to model the scheduling system. Stochastic models allow schedulers to optimize the scheduling process in the presence of randomness. For example, \cite{denton2003sequential} used a two stage stochastic model that is capable of considering flexibility in different types of costs. MDP is also a helpful stochastic dynamic programming approach to model online scheduling systems with decisions such as appointment day and time, reservation of capacity, and acceptance of patients \cite{ahmadi2017outpatient}. \cite{gocgun2014dynamic} used an MDP model to formulate a chemotherapy appointment system. In this study, we used a MDP model which decides regarding the acceptance of outpatients while the appointment day is determined based on the priority class of the outpatient who requests for appointment and the available capacity. The priority class of the outpatient is determined upon the arrival based on his/her comorbidities.
Prioritizing outpatients based on the patients characteristics and comorbidities is considered in the scheduling models to minimize the waiting time of higher priority patients in receiving appointment \cite{geng2016optimal, gocgun2011markov, patrick2008dynamic}. However, in the above studies, there is not an exact explanation and algorithm for categorizing outpatients. To the best of our knowledge, our study is the first that combines a machine learning algorithm for prioritizing outpatients with a mathematical scheduling model.

\section{Methodology}
This section is dedicated to an exploration and explanation of clustering methods and scheduling methods.

\noindent \subsection{Clustering Algorithm}
"Clustering is a data analysis technique employed to categorize data into distinct groups, to maximize the similarity of data within each group while minimizing the similarity between different groups. The primary goal is to reveal inherent patterns and structures within the dataset, facilitating a deeper understanding of the relationships and characteristics shared among data points.

\subsubsection{K-means clustering}
Partitioning clustering is the most fundamental method of cluster analysis which arranges the objects of a dataset into different exclusive clusters \cite{huang2014toward}. K-means clustering is one of the most useful partitional clustering methods. In this method, $K$ points are randomly selected from data as the centroids of the clusters. All the other points should be assigned to each cluster using the minimum distance of each point to each centroid. The centroid of each cluster is updated using the data point within it. All these iterations should be repeated until convergence criteria are met. Euclidean distance is the most useful way to calculate the distance between two points. Let $\bar{x}_i$ be the centroid of cluster $c_i$ and $d(x_j,\bar{x}_i)$ be the dissimilarity between the centroid point of each cluster and any point that belongs to that cluster (for all $x_j\in c_i$). Thus, the function to be minimized by K-means clustering can be written as follows:

\begin{align}
&\text{Min} \ \  E = \sum_{i=1}^{K}{\sum_{x_j \in c_i}^{}{d(x_j,\bar{x}_i)}}
\end{align}

%Figure \ref{fig:f1} shows the procedure of clustering data to 3 groups starting with three random initial points. 
% Algorithm \ref{al:al1} formally describes the K-means clustering approach.
% \begin{algorithm}
% %\caption{K-means partitional clustering}
% \label{al:al1}
% \begin{enumerate}
%     \item Start with K random points as centroids of clusters
%     \item Assign each point to its closest centroid  
%     \item Update the centroid of each cluster
%     \item Repeat steps 2 and 3 until the optimality criteria are met and centroids do not change
%     \end{enumerate}
% \end{algorithm}

%\medskip
%\noindent
%\begin{minipage}{\columnwidth}
%\begin{figure}
%\centering
%\includegraphics{pic1.JPEG}
%\captionof{figure}{K-means clustering procedure}\label{fig:f1}
%\end{minipage}
%\end{figure}
%\medskip
%\noindent

\subsubsection{Agglomerative hierarchical clustering}
In this method, a tree of clusters is used to separate the data, where each node represents a child cluster that is combined based on their common parent node. In hierarchical clustering, we first assign each item to a cluster such that if we have $N$ items then we would have $N$ clusters. Then, we repeat the following steps: find the closest pair of clusters and merge them into a single cluster. Compute the distance between the new cluster and each of the old clusters \cite{pendharkar2014machine}. Agglomerative clustering is one of the widely used bottom-up hierarchical methods. 
% Algorithm \ref{al:al2} shows the steps of the agglomerative clustering approach.

%\medskip
%\noindent
%\begin{minipage}{\columnwidth}
%\centering
%\includegraphics{pic2.JPEG}
%\captionof{figure}{Agglomerative clustering approach}\label{fig:f2}
%\end{minipage}
%\medskip
%\noindent

% \begin{algorithm}
% %\caption{Agglomerative hierarchical clustering}\label{al:al2}
% \begin{enumerate}
%     \item Start with n clusters
%     \item Compute the proximity matrix   
%     \item Merge the closest two clusters
%     \item Update the proximity matrix between the new cluster and the original clusters
%     \item Repeat until K clusters remains
% \end{enumerate}
% \end{algorithm}

There are different methods to combine the clusters in agglomerative clustering approaches such as Single Linkage, Complete linkage, Average linkage, Centroid method, and Ward's method \cite{defays1977efficient, el1986hierarchic, gower1969minimum, seifoddini1989single}. 

\vspace{1mm}
\subsubsection{DBSCAN clustering}
DBSCAN stands for Density-Based Spatial Clustering of Applications with Noise. It effectively clusters densely grouped data points and is particularly useful for large spatial datasets. This algorithm excels in identifying clusters of varying shapes within a dataset. To execute DBSCAN, users need to specify two essential parameters: epsilon and minPoints. Epsilon defines the radius around each data point, enabling the assessment of data density within that specific region. MinPoints denotes the minimum number of data points required within the epsilon radius for classifying a data point as a core point. In the DBSCAN context, a data point earns the Core point classification if the circle around it encompasses at least 'minPoints' data points. Points with fewer than 'minPoints' in their circle are considered Border points, and isolated points are designated as Noise. DBSCAN utilizes Euclidean distance to locate data points in space \cite{9356727}. DBSCAN operates by defining clusters as regions of high point density surrounded by regions with lower density. It initiates by choosing a random data point and expanding a cluster around it, encompassing nearby points that possess a minimum number of neighbors within a designated radius. This process is repeated until all reachable points are included in clusters, while points lacking sufficient neighbors are designated as outliers or noise.

\subsubsection{OPTICS clustering}

OPTICS, or Ordering Points To Identify the Clustering Structure, is a density-based clustering algorithm similar to DBSCAN. Unlike DBSCAN, OPTICS doesn't directly cluster the dataset but provides an ordered list based on density. It introduces reachability distance to measure local density, allowing for adaptive clustering without requiring a predefined density threshold. By identifying core points and their reachability distances, OPTICS captures nuanced density variations, offering a comprehensive understanding of the dataset's clustering structure \cite{ankerst1999optics}. OPTICS works by ordering data points based on their local densities, and reachability distances are computed; where shorter distances denote denser areas.  This method proves beneficial for datasets exhibiting diverse densities and irregular cluster shapes, as it detects clusters in the reachability plot, where consecutive points with low reachability distances coalesce into coherent groups \cite{ahmed2020analysis}.

\vspace{3mm}

In this paper, Ward's method which is an analysis of variance (ANOVA) based approach is applied. At each stage, one-way univariate ANOVAs are done for each variable with groups defined by the clusters, and two clusters that have the smallest increase in the combined error sum of squares should be merged together. 
In this paper, K-means clustering and agglomerative hierarchical clustering methods are used to group patients into two classes of high and low priority for appointment scheduling. The \cite{alizadehsani2013data} data set is used for this study. This data set contains information of 303 random patients visited Shaheed Rajaei Medical and Research Center. In this paper, for each of the patients we used 29 features that are described in Table \ref{table:t1}.

\medskip
\noindent
\begin{minipage}{\columnwidth}
\captionof{table}{Features of dataset}\label{table:t1}
\centering
\begin{tabular}{|l|c|l|c|}
\hline
Feature name & Range & Feature name & Range\\ \hline
Age & 30-86 & BP (blood pressure: mmHg) & 90-190 \\ 
Weight & 48-120 & PR (pulse rate) (ppm) & 50-110 \\ 
Length & 140-188 & Edema & Yes, No \\ 
Sex & Male, Female & Weak peripheral pulse & Yes, No \\ 
DM (history of Diabetes Mellitus) & Yes, No & Lung rales & Yes, No \\ 
HTN (history of hypertension) & Yes, No & Systolic murmur & Yes, No \\ 
Current smoker & Yes, No & Diastolic murmur & Yes, No \\ 
Ex-smoker & Yes, No & Typical Chest Pain & Yes, No \\ 
FH (history of heart disease) & Yes, No & Dyspnea & Yes, No \\ 
CRF(chronic renal failure) & Yes, No & Function class & 1,2,3,4 \\ 
CVA (Cerebrovascular Accident) & Yes, No & Atypical & Yes, No \\ 
Airway disease & Yes, No & Nonanginal CP & Yes, No \\ 
Thyroid Disease & Yes, No & Exertional CP (Exertional Chest Pain) & Yes, No \\ 
CHF (congestive heart failure) & Yes, No & Low Th Ang (low Threshold angina) & Yes, No \\
DLP (Dyslipidemia) & Yes, No & &\\ \hline
\end{tabular}
\end{minipage}   
\medskip
\noindent

To find the most important subset of features, the Wrapper method is used in this paper \cite{karegowda2010feature, talavera2005evaluation}. In this method, features are ranked based on their importance, and the best feature subset that has the best cluster quality is selected. The importance of each feature is calculated using Entropy. \cite{dash2000feature} proposed the entropy-based ranking for the first time. The entropy for each feature is calculated as follows:
\begin{align}
&E = - \sum_{i=1}^{N}{\sum_{j=1}^{N}{\Big(S_{ij}\text{log}(S_{ij})+(1-S_{ij})\text{log}(1-S_{ij})\Big)}},
\end{align}

where $S_{ij}$ is the similarity between two points $i$ and $j$, and it is calculated based on the distance between these two points after feature $t$ is removed $(dist_{i,j})$.
\begin{align}
S_{ij} = e^{-\alpha \times dist_{i,j}}
\end{align}

Here based on \cite{dash2000feature}, $\alpha$ is assumed to be $\alpha = \frac{\text{ln}(0.5)}{\bar{dist}}$  where $\bar{dist}$ is the average distance of all points after feature $t$ is removed. 
After the calculation of all features' entropy, the best features subset should be determined by calculating the cluster quality. In this study, scattering criteria are used to measure the cluster quality, considering the scatter matrix in multiple discriminant analysis. The within-cluster scatter $P_W$ and between-cluster scatter $P_B$ can be calculated in the following way:
\begin{align}
&P_W = \sum_{j=1}^{K}{\sum_{x_i \in c_j}^{}{(x_i - m_j)(x_i - m_j)^T}} \\ 
&P_B = \sum_{j=1}^{K}{(m_j - m)(m_j - m)^T}
\end{align}
Here $m$ is the total mean vector and $m_j$ is the mean vector for cluster $j$. Thus, $m$ is obtained by getting the average over all clusters, and $m_j$ is obtained by getting the average over cluster $j$. To evaluate the cluster quality using between-cluster scatter and within-cluster scatter the ``Invariant criterion'' $tr(P_W^{-1}P_B)$ is used, which measures the ratio of between-cluster to within-cluster scatter. If we add an important feature to the subset, $tr(P_W^{-1}P_B)$ increases, and if we add an unimportant feature to the subset, $tr(P_W^{-1}P_B)$ decreases or remains unchanged. 

%Figure \ref{fig:f4} shows the procedure of selecting best feature subset. 

%\medskip
%\noindent
%\begin{minipage}{\columnwidth}
%\centering
%\includegraphics{pic4.JPEG}
%\captionof{figure}{Wrapper algorithm for finding the best features subset}\label{fig:f4}
%\end{minipage}
%\medskip
%\noindent

The wrapper algorithm is one of the sequential forward feature selection algorithms to attempt the optimal feature subset by iteratively selecting features and comparing the model's performance. After ranking the features, this algorithm starts with an empty subset. At each time, one feature is added to the subset and the method performance measurement which is the ``Invariant criterion'', $tr(P^W_{-1}P_B)$ is calculated. By comparing the performance of different subsets the optimal subset can be found at each step. The procedure of adding features to subsets will continue until the stopping criteria are met. In the paper, we set the stopping criteria to be the point at which $tr(P^W_{-1}P_B)$ remains unchanged. Since the features are added to the subset based on their ranking, this algorithm calculated at maximum Q subset. 
One of the reasons that we need feature selection is that it reduces the complexity of the model, and enables the clustering method to train faster. It can also improve the accuracy of the model if the subsets are chosen correctly. Thus, we may not face over-fitting. Therefore, by using the feature selection, without losing the important information, we would have an accurate model. 
After finding the best features subset, two algorithms are applied to the data set to group the patients into multiple classes. K-means clustering and agglomerative algorithms are applied to the data set with selected features, and their accuracy is compared. Section 4 shows the result of applying these methods to the data set.

\vspace{1mm}

\subsection{Markov decision process scheduling model}\label{mdp}
As a patient requests for appointment, the system evaluates his/her health information and assigns the patient to one of the priority classes. Higher priority patients have to receive sooner appointments. Thus, based on the probable future arrivals and the priority class of the patient, the scheduler decides to give an appointment to him/her. To show how the scheduling model would work, we formulate the problem as a MDP model.

\vspace{1mm}

\subsection{Decision epochs}
At a specific point of time, when a patient requests for appointment, the scheduler observes the available capacity over an H-day booking horizon and based on the priority class of the patient, schedules him/her. Outpatients' requests for appointments arrive during the day. But in this model, we assume that the scheduler observes all the arrivals during each day, and then makes a decision about them at the end of that day. Actually, the decisions are made at the end of each day over an infinite horizon. Thus, the model becomes a discrete time MDP. In our model, the planning horizon is assumed to be rolling. It means that day $h$ at the current decision epoch becomes day $h-1$ at the next decision epoch.

\vspace{1mm}

\subsection{State space}
At the end of each day, the scheduler needs to observe the available capacity at the next $H$ days and the total number of patients that are waiting to be scheduled.
Thus, the state of the system at each decision epoch takes the form
\begin{align}
&\vec{s}=(\vec{x},\vec{y}) = (x_1,x_2,...,x_I;y_1,y_2,...,y_H),
\end{align}
Let $x_i^t$ be the number of patients with priority type $i$ on the current day (day $t$), where $i=1,2,..,I$ represents the priority class of patients. We assume that $x_i \leq D_i$ where $D_i$ shows the maximum number of type $i$ arrivals. Let $y_h$ be the number of available spots at $h$ days ahead where $h=1,2,...,H$ and $y_h \leq G$ where $G$ is the maximum capacity available at each day. In other words, $y_1 \leq y_2 \leq ... \leq y_H$ and always $y_H=G$. For this study, we choose $G=90$ and $H=7$.

\vspace{1mm}

\subsection{Action set}
The action set of the model on each day takes the form 
\begin{align}
&\vec{a}=(a_{ih}), \ \ \forall h \in \{1,2,...,H\},i \in \{1,2,...,I\},
\end{align}
where $a_{ih}$ shows the number of appointments that are offered to patients of type $i$ for $h$ days ahead. We assume that each patient needs one appointment spot. The number of offered appointments on a specific day can not exceed the number of available appointments on that day. Moreover, the number of accepted appointments on each day can not exceed the number of arrivals on that day. Thus 
\begin{align}
&\sum_{i=1}^{I}{a_{ih}} \leq y_h \ \ \forall h \in \{1,2,...,H\}, \label{a1const} \\
&\sum_{h=1}^{H}{a_{ih}} \leq x_i \ \ \forall i \in \{1,2,...,I\},\label{a2const}
\end{align}

\vspace{1mm}

\subsection{Transition probabilities}
After making a decision, the only statistical elements of the next state are new arrivals of different patient priorities. The arrivals follow Poisson distribution. Let $x'_i$ be the new arrivals of the next day. Thus, the state transition takes the form
\begin{align}
&(x_1,x_2,...,x_I;y_1,y_2,...,y_H) \rightarrow (x'_1,x'_2,...,x'_I;y_2+\sum_{i=1}^{I}{a_{i2}},y_3+\sum_{i=1}^{I}{a_{i3}},...,G),
\end{align}
which occurs with probability $p(\vec{x}')=\prod_{i=1}^I{p(x'_i)}$, where $p(x'_i)$ is the probability of arrival of $x'_i$ type $i$ patients. The demands of different priority patients are independent of each other.

\vspace{1mm}

\subsection{Costs}
The cost associated with each state action comes from two sources: \\
1. cost of giving the late appointment to type $i$ patients: $b_{ih}$, \\
2. cost of delivering or rejecting type $i$ patients: $c_{i}$, \\
Thus, the cost function takes the form
\begin{align}
c(\vec{s},\vec{a}) = \sum_{i=1}^{I}{\sum_{h=1}^{H}{b_{ih}a_{ih}}} + \sum_{i=1}^{I}{c_{i}(x_i-\sum_{h=1}^{H}{a_{ih}})},
 \end{align}
 
\subsection{The Bellman equation}
Let $v(\vec{s})$ be the total expected discounted cost over the infinite horizon. The discounting factor is denoted by $\lambda$. The Bellman equations are given by
\begin{align}
&v(\vec{s}) = \text{min} \bigg\{ c(\vec{s},\vec{a}) +\lambda \sum_{s'\in S}^{}{p(\vec{s}'|\vec{s},\vec{a})v(\vec{s}')} \bigg\} \ \ \forall \vec{s}\in S 
\label{eq:bellman}
\end{align}
The challenge is that even for very small values of arrivals, the size of the state space and the size of the action set make the problem impossible to solve by one of the traditional MDP solution methods. Thus, we refer to the fluid approximation method for solving our MDP problem. 

\vspace{1mm}

\subsection{Fluid Approximation}\label{fluidapproximation}
The state space of the MDP in Section \ref{mdp} is extremely large in practice. Therefore, due to the curse of dimensionality the classical methods for solving MDPs, e.g. value iteration, policy iteration, and linear programming techniques, cannot be used to solve the Bellman equation \eqref{eq:bellman} of the MDP model. However, we use the fluid analysis technique to approximate the MDP via a fluid model and produce sub-optimal appointment scheduling rules. Note that the fluid model is an optimal control problem. Let $x_i(t)$ be the arrival rate of priority type $i$ patients at time $t$, $y_h(t)$ be the number of available spots at $h$ days ahead at time $t$, and $u_i^h(t)$ be the rate of offering appointments to patients of type $i$ for $h$ days ahead. Then, vector $z(t)=\big(x(t),y(t)\big)$ is the vector of state variables of the optimal control problem that are our state space elements, where $x(t)=\big(x_1,\ldots,x_I\big)$ and $y(t)=\big(y_1(t),\ldots,y_H(t)\big)$ and $u(t)=\big(u_i^h(t):i\{1,\ldots,I\}, h\in \{1,\ldots,H\}\big)$ is the vector of control variables of the model. Actually, $u$ is the action vector in the optimal control model.

\vspace{1mm}

We let  $\beta=\big(\beta_i^h:i\{1,\ldots,I\}, h\in \{1,\ldots,H\}\big)$ be the late cost vector with $\beta_i^h$ being the cost associated to giving a late appointment to a patient in priority class $i$, and $\gamma=(\gamma_1,\ldots,\gamma_I)$ be the delivering/rejecting cost vector with $\gamma_i$ being the cost of delivering or rejecting a type $i$ patient. The objective function of the MDP model is the minimization of the total costs (including late appointment penalties and delivering/rejecting costs) during the planning horizon $[0,\infty)$. Thus, the objective function $F\big(z(t),u(t)\big)$ of the fluid model is given by:
\begin{align}
F\big(z(t),u(t)\big)=\int_0^{\infty}\Big(\beta u(t)+\gamma\big(x(t)-\sum_{h=1}^H u^h(t)\big)\Big)dt,
\end{align}

where $u^h(t):=\big(u_1^h(t),\ldots,u_I^h(t)\big)$. In order to write the state evolution of the system, let $\lambda(t)=\big(\lambda_1(t),\ldots,\lambda_I(t)\big)$ with $\lambda_i(t)$ being the change rate of the arrival rate of patient type $i$ at time $t$. The state variable evolution constraints can then be written as

\begin{align}
&\dot{x}_i(t)=\lambda_i(t),\ \ \  i=1,\ldots,I,\ x_i(0)=x_i^0,\label{eqfluidconst1}\\ 
&\dot{y}_h(t)=\sum_{i=1}^I u_i^{h+1}(t),\ \ \ h=1,\ldots,H,\ y_h(0)=y_h^0,\label{eqfluidconst2}
\end{align}

with the following non-negativity constraints on the state variables

\begin{align}
&x_i(t)\geq 0, \ \ \ i=1,\ldots, I,\label{eqfluidconst3}\\
&y_h(t)\geq 0,\ \ \ h=1,\ldots,H,\label{eqfluidconst4}
\end{align}

where $x_i^0$ for $i=1,\ldots,I$ and $y_h^0$ for $h=1,\ldots,H$ are the initial state of the system. Next, by using equations \eqref{a1const} and \eqref{a2const} we write the constraints on the control variables of the fluid model as follows:

\begin{align}
&\sum_{i=1}^I u_i^h(t)\leq y_h(t),\ \ \ h=1,\ldots,H,\label{eqfluidconst5}\\
&\sum_{h=1}^H u_i^h(t)\leq x_i(t),\ \ \ i=1,\ldots,I,\label{eqfluidconst6}
\end{align}

with the following non-negativity constraints on the control variables: 

\begin{align}
u_i^h(t)\geq 0, \ \ \ i=1,\ldots,I,\ h=1,\ldots,H.\label{eqfluidconst7}
\end{align}

Then, the set of feasible actions of the fluid model is given by 

\begin{align}
\Omega(t)=\Big\{&u(t):\sum_{i=1}^I u_i^h(t)\leq y_h(t),\ h=1,\ldots,H\ ; \sum_{h=1}^H u_i^h(t)\leq x_i(t),\ i=1,\ldots,I;\nonumber\\
& u_i^h(t)\geq 0,\ i=1,\ldots,I,\ h=1,\ldots,H \Big\}.
\end{align}

Hence, the fluid approximation of the stochastic MDP formulation of the appointment scheduling system is as follows

\begin{equation}
\tag{$P_1$}
\begin{cases}
&V_F(z_0)=\max \int_0^{\infty}\Big(\beta u(t)+\gamma\big(x(t)-\sum_{h=1}^H u^h(t)\big)\Big)dt\\
&\text{subject to }(\ref{eqfluidconst1})-(\ref{eqfluidconst7}),
\end{cases}
\end{equation}

which is a linear optimal control problem with mixed state-action constraints as the set of feasible control $\Omega(t)$ includes constraints on both state and control variables.

\vspace{1mm}

We use the necessary and sufficient conditions for the optimality of a solution to the optimal control problem $(P_1)$, which is discussed in \cite{hartl1995survey}, to characterize the structure of its optimal solution. 

\vspace{1mm}

\section{Numerical results}
The clustering methods implemented in this paper include K-mean, Agglomerative hierarchical, DBSCAN, and OPTICS. The Wrapper algorithm is used to determine the most suitable subset of features prior to applying the clustering methods.
To determine the optimal number of clusters for K-means and Agglomerative Hierarchical methods, we utilized the elbow method.  problem calculated using the elbow method. In this approach, the K-means clustering method is applied for each number of cluster k between 1 and 10, and the sum of squared errors (SSE) is calculated. SSE will always decrease with larger K. (SSE is 0 when K is equal to the number of data points in the data set, because then each data point is its cluster, and there is no error between it and the center of its cluster). At some point, the SSE will not decrease much between values, this implies that probably two centers are used in the same grouping of data, so the squared distance to either is similar. By employing the elbow method and examining the plot of Within-Cluster Sums of Squares (WCSS) against k, we can identify the optimal cluster configuration for addressing the problem.
DBSCAN and OPTICS do not rely on a predefined number of clusters like k-means does. The number of clusters is identified based on the density of data points rather than assuming a specific number of clusters.
\medskip
\noindent
\begin{minipage}{\columnwidth}
\centering
\includegraphics[width=6.5cm,height=4.5cm]{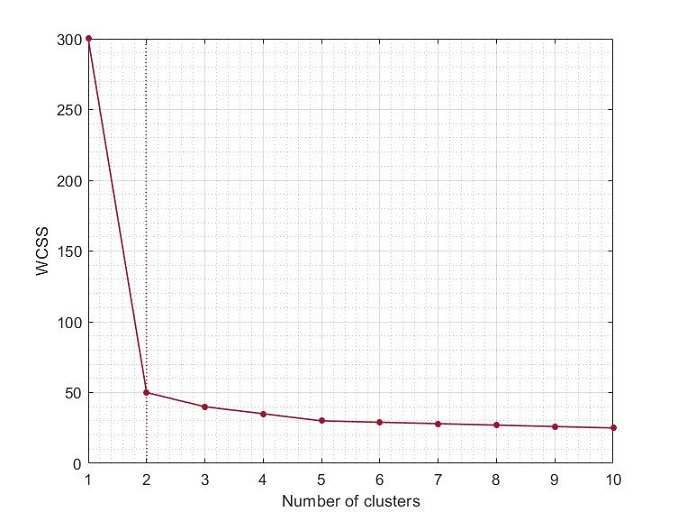}
\captionof{figure}{Elbow method for optimal number of clusters}\label{fig:f5}
\end{minipage}
\medskip
\noindent

As shown in Figure \ref{fig:f5}, after K=2, the rate of decrease in the sum of squared distances slows down, suggesting that the optimal number of clusters is K=2. Additionally, for DBSCAN and OPTICS, a similar number of clusters (K=2) is determined based on the density of the data set.
The proposed clustering methods are applied to group the patients into two different classes of high and low priority. 
To compare the accuracy of these clustering methods, the Silhouette coefficient is utilized. The Silhouette coefficient measures the similarity of a point to its own cluster relative to other clusters. It can be computed as follows:
\begin{align}
&sil(i) = \frac{\mu(i) - \zeta(i)}{\text{Max} \{ \zeta(i),\mu(i)\}},
\end{align}

where $\zeta(i)$ is the average distance of point $i$ from all the other points in the same cluster $(C_i)$, and $\mu(i)$ is smallest average distance of point $i$ to all points in other clusters. $\mu(i)=\text{Min}{d(i,C)}$, where $d(i,C)$ is the average distance of point $i$ to other points in cluster $C \neq C_i$. Comparing the average silhouette coefficient for all four methods, it can be concluded that DBSCAN clustering has better performance with a silhouette score of 0.87.
Table \ref{table:score} shows the silhouette score for all four methods.

\medskip
\noindent
\begin{minipage}{\columnwidth}
\captionof{table}{The silhouette score for clustering methods.}\label{table:score}
\centering
\begin{tabular}{|c|c|c|c|c|}
\hline
   & K-means & Agglomerative Hierarchical & DBSCAN & OPTICS \\ \hline
Score & 0.87 & 0.81 & 0.92 & 0.89 \\ 
 \hline
\end{tabular}
\end{minipage}   
\medskip
\noindent

This paper recommends employing K-means clustering as a decision-making tool for classifying incoming patients into priority classes before scheduling appointments. Grouping patients beforehand allows for more efficient and targeted scheduling, optimizing resource allocation, and improving overall system efficiency. 

Using the data set, the daily arrivals of outpatient types are estimated as Poisson distributions with means $\lambda_1 =44$ and $\lambda_2 =56$, where $\lambda_1$ and $\lambda_2$ represent the arrival rates of lower and higher priority outpatients, respectively. Outpatients are scheduled within a week (H=7). According to the priority classes we estimate the cost parameters of the MDP model. The estimations are provided in Table \ref{table:parametersestimation}. According to the table, $b_{ih}$ is the cost of giving an appointment $h$ days later to outpatient class $i$, and $c_{i}$ is the cost of rejecting an outpatient of priority class {i}. $G$ is the total number of available appointments each day. 

\medskip
\noindent
\begin{minipage}{\columnwidth}
\captionof{table}{Parameters values for the MDP model}\label{table:parametersestimation}
\centering
\begin{tabular}{|c|c||c|c|}
\hline
 Parameter& Value & Parameter & Value  \\ \hline
$b_{11}$ & 0 & $b_{23}$ & 3  \\ 
$b_{12}$ & 0.5 & $b_{24}$ & 0  \\
$b_{13}$ & 1 & $b_{25}$ & 1   \\ 
$b_{14}$ & 1.5 & $b_{26}$ & 2   \\ 
$b_{15}$ & 2 & $b_{27}$ & 3  \\ 
$b_{16}$& 2.5 & $c_{1}$ & 5 \\ 
$b_{17}$ & 3 & $c_{2}$ & 10  \\ 
$b_{21}$ & 0 & $G$ & 90 \\ 
$b_{22}$ & 1 &  &  \\ 
 \hline
\end{tabular}
\end{minipage}   
\medskip
\noindent

After solving the MDP using the fluid approximation method the optimal policy is obtained. According to the optimal policy, the optimal number of appointments that have to be scheduled within the next 7 days is estimated. The results are represented in Table \ref{table:result}. The remaining appointment requests have to be canceled. In other words, after observing the optimal policy we come up with a benchmark policy that is easy to implement in the healthcare facility. Using this benchmark policy that is obtained based on the optimal policy, the healthcare facility does not need to solve the MDP every day. The scheduler just needs to classify the arrivals and then use the benchmark policy for scheduling them. According to this benchmark policy on each day, the number of outpatients of each group that we schedule on the next 7 days is based on the values of Table \ref{table:result}. For instance, at the end of each day, we schedule 15 outpatients of class 1 and 20 outpatients of class 2 to the next day appointments. For two days ahead, we always schedule 12 and 17 outpatients of class 1 and 2, respectively. We continue this procedure for the next 7 days. Therefore, at the end of each day, we schedule some outpatients for the next 7 days. Moreover, we should mention that after fulfilling the appointments based on the proposed numbers, if we still have available appointments for the next day, we schedule higher priority outpatients to make sure that we use all the capacity. In the other case, if we do not have enough capacity as much as the individuals we are supposed to schedule for the next day, we decrease the number of lower priority outpatients that are scheduled for the next day. These cases would happen because of the uncertain arrivals in the real world.

\medskip
\noindent
\begin{minipage}{\columnwidth}
\captionof{table}{Number of appointments that can be scheduled based on the optimal policy}\label{table:result}
\centering
\begin{tabular}{|c|c|c|c|c|c|c|c|}
\hline
 Priority class & day 1 & day 2 & day 3 & day 4 & day 5 & day 6 & day 7  \\ \hline
1 & 15 & 12 & 9 & 6 & 3 & 0 & 0 \\ 
2 & 20 & 17 & 14 & 11 & 8 & 5 & 2 \\
 \hline
\end{tabular}
\end{minipage}   
\medskip
\noindent

\vspace{1mm}

\section{Conclusion}
In this paper, we use K-means clustering, agglomerative hierarchical clustering, DBSCAN, and OPTICS to categorize outpatients into high and low priority classes. The wrapper algorithm is used to find the best feature of the data set to be used in training the clustering pattern. The accuracy of the clustering methods is compared using the Silhouette coefficient, revealing that DBSCAN clustering exhibits higher accuracy than other methods. Consequently, the DBSCAN method is recommended for constructing a pattern and predicting the priority classes of outpatients. Subsequently, the outpatient scheduling model is applied for appointment scheduling. The scheduling process is formulated as a Markov Decision Process (MDP), and a fluid approximation is employed for solving it. The model aims to provide the most immediate appointments for higher-priority outpatients. Building on the outcomes of the MDP, we introduce a benchmark policy for efficiently scheduling multiple classes of outpatients. We used \cite{alizadehsani2013data} data set and applied our scheduling procedure.

\vspace{1.5cm}

\small
\noindent
%\textit{\textbf{Nooshin Yousefi}
%is currently a PhD student at Rutgers university, department of industrial engineering. She got her master degree from West Virginia University.}\\
%\includegraphics[height=10pt, width=10pt]{orcid_128x128.eps} %\url{https://orcid.org/0000-0001-XXXX-XXXX}

\vspace{1.5cm}
%\noindent
%\textit{\textbf{Mahsa Kiani}
%is currently a PhD student at Clemson university, department of industrial engineering. She got her master %degree from West Virginia University.}\\
%\includegraphics[height=10pt, %width=10pt]{orcid_128x128.eps} %\url{https://orcid.org/0000-0001-XXXX-XXXX}

\end{document}